\documentclass[aps,prd,twocolumn,superscriptaddress]{revtex4}
\usepackage{epsfig,epsf}
\usepackage{amsmath}
\usepackage{amsthm}
\usepackage{amsfonts}
\usepackage{amssymb}
\usepackage{dsfont}
\usepackage{multirow}
\usepackage{appendix}
\usepackage{slashed}
\usepackage[active]{srcltx}
\usepackage{psfrag}

\begin{document}

\title{Axial-vector molecules $\Upsilon B_{c}^{-}$ and $\eta_{b}B_{c}^{%
\ast-} $}
\date{\today}
\author{S.~S.~Agaev}
\affiliation{Institute for Physical Problems, Baku State University, Az--1148 Baku,
Azerbaijan}
\author{K.~Azizi}
\thanks{Corresponding Author: kazem.azizi@ut.ac.ir}
\affiliation{Department of Physics, University of Tehran, North Karegar Avenue, Tehran
14395-547, Iran}
\affiliation{Department of Physics, Dogus University, Dudullu-\"{U}mraniye, 34775
Istanbul, T\"{u}rkiye}
\author{H.~Sundu}
\affiliation{Department of Physics Engineering, Istanbul Medeniyet University, 34700
Istanbul, T\"{u}rkiye}

\begin{abstract}
Axial-vector hadronic molecules $\mathcal{M}_{\mathrm{AV}}=\Upsilon
B_{c}^{-} $ and $\widetilde{\mathcal{M}}_{\mathrm{AV}}=\eta_{b}B_{c}^{\ast
-} $ with the quark content $bb \overline{b}\overline{c}$ are studied using
QCD sum rule method. The spectroscopic parameters of these molecules are
computed in the context of the two-point sum rule method. Predictions for
their masses are identical to each other and confirm that they are
structures unstable against dissociations to ordinary heavy mesons. We
evaluate the width of the state $\mathcal{M}_{\mathrm{AV}}$ and assume that
it is equal to that of $\widetilde{\mathcal{M}}_{\mathrm{AV}} $. To this
end, we explore its dominant decay channels $\mathcal{M}_{\mathrm{AV}} \to
\Upsilon B_{c}^{-} $ and $\mathcal{M}_{\mathrm{AV}} \to \eta_{b}B_{c}^{\ast
-}$. There also are subleading modes of $\mathcal{M}_{\mathrm{AV}}$
generated due to annihilation of $\overline{b}b$ quarks. We consider decays
of the molecule $\mathcal{M}_{\mathrm{AV}}$ to pairs of the mesons $B^{\ast
-} \overline{D}^{0}$, $\overline{B}^{\ast 0} D^{-}$, $B^{-} \overline{D}
^{\ast 0}$, $\overline{B}^{0} D^{\ast -}$, $\overline{B}_{s}^{\ast 0}
D_{s}^{-}$, and $\overline{B}_{s}^{0} D_{s}^{\ast -}$. To find strong
couplings at the $\mathcal{M}_{\mathrm{AV}}$-meson-meson vertices which
determine the partial widths of these processes, we apply QCD three-point
sum rule approach. The mass $m=(15800 \pm 90)~\mathrm{MeV}$ and width $%
\Gamma [\mathcal{M}_{\mathrm{AV}}]=(114 \pm 17)~ \mathrm{MeV}$ of the
molecule $\mathcal{M}_{\mathrm{AV}}$ are useful for experimental studies of
fully heavy molecular structures at ongoing and planning experiments.
\end{abstract}

\maketitle


\section{Introduction}

\label{sec:Intro}

It is known that hadronic molecular structures are among objects interesting
for high energy physics. These exotic states are presumably composed of
ordinary mesons and establish a class in the hadron spectroscopy. Though
existence of different hadronic molecules was already supposed \cite%
{Bander:1975fb,Voloshin:1976ap,DeRujula:1976zlg}, binding mechanisms of
these structures, the processes at which they might be discovered and
methods for computations of the parameters of such states were elaborated
mainly in the later publications \cite%
{Tornqvist:1991lks,Ding:2008mp,Zhang:2009vs,Sun:2012sy,Chen:2015ata,Karliner:2015ina, Liu:2016kqx,Chen:2017vai,Sun:2018zqs,Albuquerque:2012rq,PavonValderrama:2019ixb,Molina:2020hde,Xu:2020evn, Xin:2021wcr,Agaev:2022duz,Agaev:2023eyk,Braaten:2023vgs,Wu:2023rrp,Liang:2023jxh,Wang:2025zss,Braaten:2024tbm}%
.

States that are built of only heavy mesons form an interesting and rapidly
growing subclass of the hadronic molecules. They may contain an equal number
of charm or bottom quarks-antiquarks. This group of molecules was studied in
Refs.\ \cite{Agaev:2023ruu,Agaev:2023rpj,Yalikun:2025ssz,Liu:2024pio}. We
also explored the fully heavy molecules $\eta _{c}\eta _{c}$, $\chi
_{c0}\chi _{c0}$, and $\chi _{c1}\chi _{c1}$ by calculating their masses and
decay widths \cite{Agaev:2023ruu,Agaev:2023rpj}. In these articles we
suggested their interpretation as possible candidates of new four $X$
resonances which supposedly are $cc\overline{c}\overline{c}$ particles \cite%
{LHCb:2020bwg,Bouhova-Thacker:2022vnt,CMS:2023owd}.

Hidden charm-bottom molecules belong to this group of particles as well, and
were investigated in Refs.\ \cite%
{Liu:2024pio,Liu:2023gla,Wang:2023bek,Agaev:2025wdj,Agaev:2025fwm,Agaev:2025nkw}%
. Thus, in the framework of the coupled-channel unitary approach\ the
properties of molecules $B_{c}^{(\ast )+}B_{c}^{(\ast )-}$ were addressed in
Ref. \cite{Liu:2023gla}. The parameters and decay channels of the states $%
B_{c}^{+}B_{c}^{-}$, $(B_{c}^{\ast +}B_{c}^{-}+B_{c}^{+}B_{c}^{\ast -})/2$
and $B_{c}^{\ast +}B_{c}^{\ast -}$ with spin-parities $J^{\mathrm{PC}%
}=0^{++} $, $1^{++}$, and $2^{++}$ were explored in our publications \cite%
{Agaev:2025wdj,Agaev:2025fwm,Agaev:2025nkw}.

Hadronic molecules with the nonsymmetrical contents $bb\overline{b}\overline{%
c}$ and $cc\overline{c}\overline{b}$ establish another group of fully heavy
particles. It is worth noting that, information about these structures is
rather limited. Thus, these states are yet not observed in experiments.
There are a few publications devoted to theoretical analyses of such
molecules \cite{Liu:2024pio,Agaev:2025did}. In our work \cite{Agaev:2025did}
we calculated the mass and decay width of the scalar states $\mathcal{M}_{%
\mathrm{b}}=\eta _{b}B_{c}^{-}$ and $\mathcal{M}_{\mathrm{c}}=\eta
_{c}B_{c}^{+}$: Our results demonstrated that they are relatively broad
compounds. But at lower limit of its mass $\mathcal{M}_{\mathrm{b}}$ may
form a bound state of mesons $\eta _{b}$ and $B_{c}^{-}$. These interesting
facts make the molecules with the contents $bb\overline{b}\overline{c}$ and $%
cc\overline{c}\overline{b}$ attractive objects for researches.

In this article, we investigate the axial-vector hadronic molecules $%
\mathcal{M}_{\mathrm{AV}}=\Upsilon B_{c}^{-}$ and $\widetilde{\mathcal{M}}_{%
\mathrm{AV}}=\eta _{b}B_{c}^{\ast -}$ and calculate their masses and full
decay widths. These structures composed of $bb\overline{b}\overline{c}$
quarks are counterparts of the scalar molecule $\mathcal{M}_{\mathrm{b}}$
from our previous paper \cite{Agaev:2025did}. Here, computations are
performed in the context of QCD SR method \cite%
{Shifman:1978bx,Shifman:1978by}. Parameters of $\mathcal{M}_{\mathrm{AV}}$
and $\widetilde{\mathcal{M}}_{\mathrm{AV}}$ show that despite different
internal organizations, they have almost identical masses: There is a
difference around of a few $\mathrm{MeV}$ which is considerably smaller than
the accuracy of the sum rule method. Therefore, we treat the molecules $%
\mathcal{M}_{\mathrm{AV}}$ and $\widetilde{\mathcal{M}}_{\mathrm{AV}}$ as
identical particles, and concentrate on properties of $\mathcal{M}_{\mathrm{%
AV}}$.

The structure $\mathcal{M}_{\mathrm{AV}}$ is strong-interaction unstable
particle and easily dissociates to constituent mesons $\Upsilon B_{c}^{-}$.
The decay to a pair $\eta _{b}B_{c}^{\ast -}$ is also among its
kinematically allowed decay channels. These two processes are dominant decay
modes of $\mathcal{M}_{\mathrm{AV}}$, in which all quarks from $\mathcal{M}_{%
\mathrm{AV}}$ participate in formation of \ the final-state mesons with
required spin-parities. \ Besides, there are channels which are triggered by
annihilation of $b\overline{b}$ quarks and generation of light
quark-antiquark pairs $\overline{q}q$ and $\overline{s}s$. Having combined
with the remaining heavy quarks they create conventional meson pairs. In
this work, we explore production of mesons $B^{\ast -}\overline{D}^{0}$, $%
\overline{B}^{\ast 0}D^{-}$, $B^{-}\overline{D}^{\ast 0}$,$\overline{B}%
^{0}D^{\ast -}$, $\overline{B}_{s}^{\ast 0}D_{s}^{-}$, and $\overline{B}%
_{s}^{0}D_{s}^{\ast -}$ due to this mechanism. This mechanism for
transformation of four-quark states was considered in Refs.\ \cite%
{Becchi:2020mjz,Becchi:2020uvq,Agaev:2023ara} and applied to decays of
various diquark-antidiquark states. It was successfully used to study
processes with molecular structures as well \cite%
{Agaev:2025wdj,Agaev:2025fwm,Agaev:2025nkw,Agaev:2025did}.

Partial widths all of these channels depend on numerous parameters such as
masses and current couplings (decay constants) of particles involved into a
process. They contain also strong couplings of particles at the relevant
molecule-meson-meson vertices. These couplings are objects of our present
calculations. They can be estimated by applying technical tools of QCD
three-point SR method. But there are differences in treatments of the
dominant and subleading modes. Thus, in the case of the dominant channels,
we apply the standard techniques to compute a correlation function of
interest, whereas the subleading decays require additionally a replacement
the vacuum expectation value $\langle \overline{b}b\rangle $ by the gluon
condensate $\langle \alpha _{s}G^{2}/\pi \rangle $. Nevertheless this
operation does not imply introduction into analysis of new parameters.

This paper is structured in the following from: In Sec.\ \ref{sec:Mass}, we
calculate the spectroscopic parameters of the axial-vector molecules $%
\mathcal{M}_{\mathrm{AV}}$ and $\widetilde{\mathcal{M}}_{\mathrm{AV}}$. The
Sec.\ \ref{sec:Widths1} is devoted to investigation of the dominant decay
channels $\mathcal{M}_{\mathrm{AV}}\rightarrow \Upsilon B_{c}^{-}$ and $%
\mathcal{M}_{\mathrm{AV}}\rightarrow \eta _{b}B_{c}^{\ast -}$. The
subleading decays of the molecule $\mathcal{M}_{\mathrm{AV}}$ are considered
in Sec. \ref{sec:Widths2}. Here, we also find the full decay width of $%
\mathcal{M}_{\mathrm{AV}}$. The last section \ref{sec:Conc} contains our
analysis of the results and a few final notes.


\section{Spectroscopic parameters of the molecules $\mathcal{M}_{\mathrm{AV}%
} $ and $\widetilde{\mathcal{M}}_{\mathrm{AV}}$}

\label{sec:Mass}

To evaluate the masses and current couplings (pole residues) of the
molecular structures $\mathcal{M}_{\mathrm{AV}}$ and $\widetilde{\mathcal{M}}%
_{\mathrm{AV}}$, we introduce the relevant interpolating currents%
\begin{equation}
J_{\mu }(x)=\overline{b}_{a}(x)\gamma _{\mu }b_{a}(x)\overline{c}%
_{b}(x)i\gamma _{5}b_{b}(x),
\end{equation}%
and
\begin{equation}
\widetilde{J}_{\mu }(x)=\overline{b}_{a}(x)i\gamma _{5}b_{a}(x)\overline{c}%
_{b}(x)\gamma _{\mu }b_{b}(x).
\end{equation}%
Here, $b(x)$ and $c(x)$ are the quark fields, whereas $a$ and $b$ denote the
color indices.

The SRs for the mass $m$ and current coupling $\Lambda $ of $\mathcal{M}_{%
\mathrm{AV}}$ can be extracted from analysis of the following correlation
function
\begin{equation}
\Pi _{\mu \nu }(p)=i\int d^{4}xe^{ipx}\langle 0|\mathcal{T}\{J_{\mu
}(x)J_{\nu }^{\dag }(0)\}|0\rangle ,  \label{eq:CF1}
\end{equation}%
with $\mathcal{T}$ \ being the time-ordered product of two currents.

In the framework of the sum rule method the correlator $\Pi _{\mu \nu }(p)$
should be calculated by two alternative ways. Thus, it has to be found using
the parameters $m$ and $\Lambda $. The correlation function calculated by
this manner forms the physical side $\Pi _{\mu \nu }^{\mathrm{Phys}}(p)$ of
the corresponding SRs. The correlator $\Pi _{\mu \nu }^{\mathrm{Phys}}(p)$
is determined by the expression
\begin{eqnarray}
\Pi _{\mu \nu }^{\mathrm{Phys}}(p) &=&\frac{\langle 0|J_{\mu }|\mathcal{M}_{%
\mathrm{AV}}(p,\epsilon )\rangle \langle \mathcal{M}_{\mathrm{AV}%
}(p,\epsilon )|J_{\nu }^{\dagger }|0\rangle }{m^{2}-p^{2}}  \notag \\
&&+\cdots ,  \label{eq:Phys1}
\end{eqnarray}%
where $\epsilon _{\mu }$ is the polarization vector of the axial-vector
molecule. The term presented above explicitly is the contribution of the
ground-level molecule, whereas ellipses denote contributions of the higher
resonances and continuum states.

We calculate $\Pi _{\mu \nu }^{\mathrm{Phys}}(p)$ by introducing the matrix
element
\begin{equation}
\langle 0|J_{\mu }|\mathcal{M}_{\mathrm{AV}}(p,\epsilon )\rangle =\Lambda
\epsilon _{\mathcal{\mu }}(p),  \label{eq:ME1}
\end{equation}%
and obtain
\begin{equation}
\Pi _{\mu \nu }^{\mathrm{Phys}}(p)=\frac{\Lambda ^{2}}{m^{2}-p^{2}}\left(
g_{\mu \nu }-\frac{p_{\mu }p_{\nu }}{m^{2}}\right) +\cdots .
\label{eq:Phys2}
\end{equation}%
The correlation function contains two Lorentz structures. Because the term $%
\sim g_{\mu \nu }$ receives contributions from the spin-1 particle, we
choose it to carry our the sum rule analysis. Then the factor $\Lambda
^{2}/(m^{2}-p^{2})$ is the invariant amplitude $\Pi ^{\mathrm{Phys}}(p^{2})$
necessary for future studies.

The correlator $\Pi _{\mu \nu }(p)$ should be computed by employing heavy
quark propagators $S_{b(c)}(x)$ and using techniques of the operator product
expansion ($\mathrm{OPE}$) . As a result, we get
\begin{eqnarray}
&&\Pi _{\mu \nu }^{\mathrm{OPE}}(p)=i\int d^{4}xe^{ipx}\left\{ \mathrm{Tr}%
\left[ \gamma _{\mu }S_{b}^{ab^{\prime }}(x)\gamma _{5}S_{c}^{b^{\prime
}b}(-x)\gamma _{5}\right. \right.  \notag \\
&&\left. \times S_{b}^{ba^{\prime }}(x)\gamma _{\nu }S_{b}^{a^{\prime
}a}(-x) \right] -\mathrm{Tr}\left[ \gamma _{\mu }S_{b}^{aa^{\prime
}}(x)\gamma _{\nu }S_{b}^{a^{\prime }a}(-x)\right]  \notag \\
&&\left. \times \mathrm{Tr}\left[ \gamma _{5}S_{b}^{bb^{\prime }}(x)\gamma
_{5}S_{c}^{b^{\prime }b}(-x)\right] \right\} .  \label{eq:QCD1}
\end{eqnarray}%
The explicit expressions for $S_{b(c)}(x)$ can be found in Ref. \cite%
{Agaev:2020zad}. The function $\Pi _{\mu \nu }^{\mathrm{OPE}}(p)$
establishes the QCD component of the sum rules.

The function $\Pi _{\mu \nu }^{\mathrm{OPE}}(p)$ is also composed of two
terms. We choose a term which is proportional to $g_{\mu \nu }$ and label by
$\Pi ^{\mathrm{OPE}}(p^{2})$ corresponding invariant amplitude. Afterwards,
we equate amplitudes $\Pi ^{\mathrm{Phys}}(p^{2})$ and $\Pi ^{\mathrm{OPE}%
}(p^{2})$, carry out required operations detailed in Ref. \cite%
{Agaev:2024uza}, and obtain the sum rules for parameters $m$ and $\Lambda $
of the molecule $\mathcal{M}_{\mathrm{AV}}$
\begin{equation}
m^{2}=\frac{\Pi ^{\prime }(M^{2},s_{0})}{\Pi (M^{2},s_{0})},  \label{eq:Mass}
\end{equation}%
and
\begin{equation}
\Lambda ^{2}=e^{m^{2}/M^{2}}\Pi (M^{2},s_{0}).  \label{eq:Coupl}
\end{equation}

Above, we use the notation $\Pi ^{\prime }(M^{2},s_{0})=d\Pi
(M^{2},s_{0})/d(-1/M^{2})$, where $\Pi (M^{2},s_{0})$ is the amplitude $\Pi
^{\mathrm{OPE}}(p^{2})$ subjected to Borel transformation and continuum
subtraction procedures. The first of them is done to subdue effects of
higher resonances and continuum states, whereas the second procedure permits
one to subtract these terms from the QCD side of the SR formula. Then, $\Pi
(M^{2},s_{0})$ becomes a function of the Borel $M^{2}$ and continuum
subtraction $s_{0}$ parameters. In the case under analysis, it is given by
the formula%
\begin{equation}
\Pi (M^{2},s_{0})=\int_{(3m_{b}+m_{c})^{2}}^{s_{0}}ds\rho ^{\mathrm{OPE}%
}(s)e^{-s/M^{2}}+\Pi (M^{2}).  \label{eq:CorrF}
\end{equation}%
Here, $\rho ^{\mathrm{OPE}}(s)$ is the spectral density found as an
imaginary part of the function $\Pi ^{\mathrm{OPE}}(p^{2})$. In the current
work we take into account contributions to $\Pi ^{\mathrm{OPE}}(p^{2})$
coming from the perturbative and dimension-four terms $\sim \langle \alpha
_{s}G^{2}/\pi \rangle $. The reason is that dimension-$6$ contributions
which are proportional to the triple-gluon condensate $\langle
g_{s}^{3}G^{3}\rangle $ in the case of the heavy hadronic molecules are
negligibly small: This was demonstrated by explicit calculations in Ref.\
\cite{Agaev:2025did} when exploring the scalar molecule $\eta _{b}B_{c}^{-}$%
. Therefore, we truncate $\mathrm{OPE}$ at $\langle \alpha _{s}G^{2}/\pi
\rangle $ level and neglect higher-dimension terms. As a result, $\rho ^{%
\mathrm{OPE}}(s)$ consists of $\rho ^{\mathrm{pert.}}(s)$ and $\rho ^{%
\mathrm{Dim4}}(s)$ components. The function $\Pi (M^{2})$ in Eq.\ (\ref%
{eq:CorrF}) is extracted from the correlation function $\Pi ^{\mathrm{OPE}%
}(p)$ and does not contain terms included into $\rho ^{\mathrm{OPE}}(s)$.

To perform computations, one should specify the parameters in Eqs.\ (\ref%
{eq:Mass}) and (\ref{eq:Coupl}). The gluon condensate $\langle \alpha
_{s}G^{2}/\pi \rangle =(0.012\pm 0.004)~\mathrm{GeV}^{4}$ as well as masses $%
m_{b}=(4.183\pm 0.007)~\mathrm{GeV}$ and $m_{c}=(1.2730\pm 0.0046)~\mathrm{%
GeV}$ of the quarks are universal parameters \cite%
{Shifman:1978bx,Shifman:1978by,PDG:2024}. Contrary, the pair $M^{2}$ and $%
s_{0}$ is fixed by a problem under consideration and should meet well-known
constraints of SR investigations. One of them is dominance of the pole
contribution ($\mathrm{PC}$) to obtained quantities: As a result, one
demands fulfilment of the restriction $\mathrm{PC}\geq 0.5$ . Convergence of
$\mathrm{OPE}$ is the next constraint of credible SR investigations.
Because, the correlator $\Pi (M^{2},s_{0})$ contains dimension-$4$
nonperturbative term $\Pi ^{\mathrm{Dim4}}(M^{2},s_{0})$, it is enough
fulfillment of $|\Pi ^{\mathrm{Dim4}}(M^{2},s_{0})|\leq 0.05|\Pi
(M^{2},s_{0})|$ which guarantees convergence of $\mathrm{OPE}$. One has also
to remember that extracted quantities should be stable against variations of
$M^{2}$ and $s_{0}$.

\begin{figure}[h]
\includegraphics[width=8.5cm]{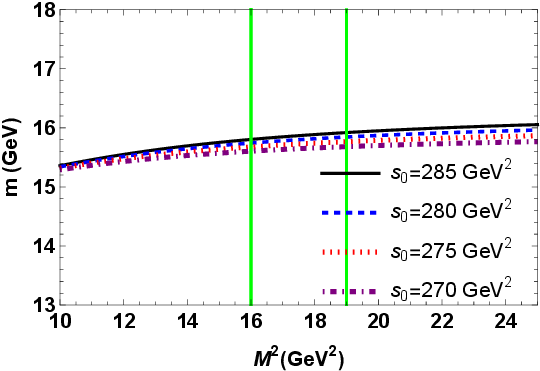}
\caption{The mass $m$ as a function of parameters $M^{2}$ at various $s_{0}$%
. Two vertical lines fix borders of $M^{2}$ inside of which restrictions
imposed on $\Pi (M^{2},s_{0})$ are fulfilled. }
\label{fig:WR}
\end{figure}

Armed with this knowledge, we have performed numerical computations of $m$
over wide range of the parameters $M^{2}$ and $s_{0}$: Some of these
computations are visualized in Fig.\ \ref{fig:WR}.

\begin{figure}[h]
\includegraphics[width=8.5cm]{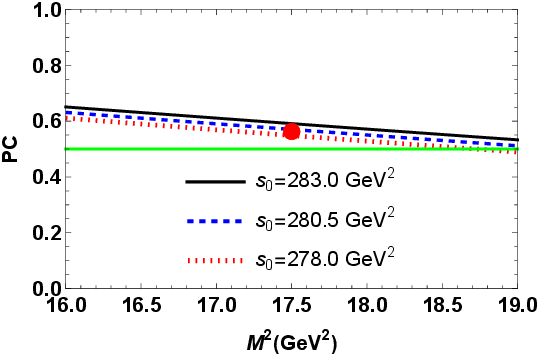}
\caption{Dependence of $\mathrm{PC}$ on the Borel parameter $M^{2}$ at fixed
$s_{0}$. The circle shows the point $M^{2}=17.5~\mathrm{GeV}^{2}$ and $%
s_{0}=280.5~\mathrm{GeV}^{2}$. }
\label{fig:PC}
\end{figure}

Collected predictions allow us to find the windows for $M^{2}$ and $s_{0}$,
where all restrictions discussed above are satisfied. We find that intervals
\begin{equation}
M^{2}\in \lbrack 16,19]~\mathrm{GeV}^{2},\ s_{0}\in \lbrack 278,283]~\mathrm{%
GeV}^{2},  \label{eq:Wind1}
\end{equation}%
meet these constraints. Indeed, at maximal $M^{2}$ the $s_{0}$-averaged pole
contribution is $\mathrm{PC}\approx 0.51$, whereas at minimal value of the
Borel parameter it is equal to $\mathrm{PC}\approx 0.63$. At $M^{2}=16~%
\mathrm{GeV}^{2}$ the nonperturbative term is negative and forms
approximately $1.4\%$ of $\Pi (M^{2},s_{0})$. As a function of $M^{2}$ the $%
\mathrm{PC}$ is depicted in Fig.\ \ref{fig:PC}, where all lines exceed the
limit $\mathrm{PC}=0.5$.

We extract $m$ and $\Lambda $ as mean values of these parameters in the
regions Eq.\ (\ref{eq:Wind1}) and obtain
\begin{eqnarray}
m &=&(15800\pm 90)~\mathrm{MeV},  \notag \\
\Lambda &=&(3.33\pm 0.35)~\mathrm{GeV}^{5}.  \label{eq:Result1}
\end{eqnarray}%
The results in Eq.\ (\ref{eq:Result1}) are effectively equal to SR
predictions at $M^{2}=17.5~\mathrm{GeV}^{2}$ and $s_{0}=280.5~\mathrm{GeV}%
^{2}$, where $\mathrm{PC}\approx 0.57$ ensuring the dominance of $\mathrm{PC}
$ in $m$ and $\Lambda $. The uncertainties in Eq.\ (\ref{eq:Result1}) are
equal to $\pm 0.6\%$ for the mass $m$, and to $\pm 11\%$ \ for the current
coupling $\Lambda $. They appear mainly due to choices of $M^{2}$ and $s_{0}$%
: Ambiguities generated by errors in quark masses and gluon condensate are
negligibly small. Note that these theoretical errors remain within limits
accepted in SR analyses.

\begin{widetext}

\begin{figure}[h!]
\begin{center}
\includegraphics[totalheight=6cm,width=8cm]{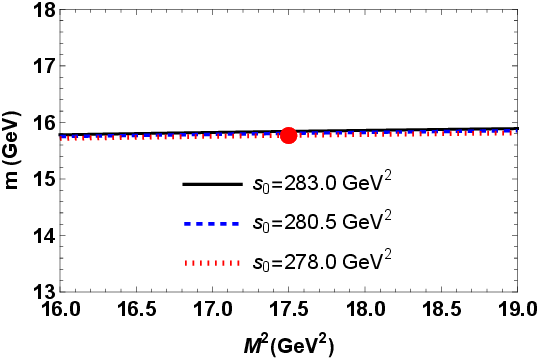}
\includegraphics[totalheight=6cm,width=8cm]{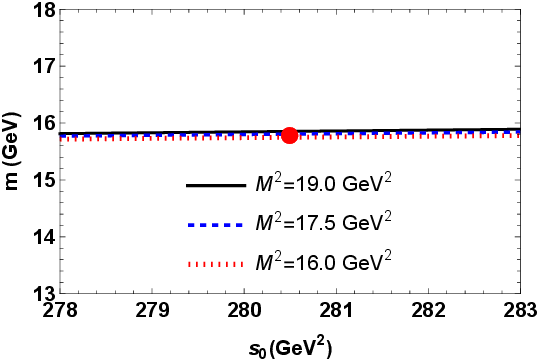}
\end{center}
\caption{The mass $m$ as a function of $M^{2}$ (left panel), and $s_0$ (right panel).}
\label{fig:Mass}
\end{figure}

\end{widetext}

Dependence of $m$ on the parameters $M^{2}$ and $s_{0}$ is shown in Fig.\ %
\ref{fig:Mass}. Having inspected curves presented there one can be convinced
in stability of this prediction.

Investigations aimed to find the parameters $\widetilde{m}$ and $\widetilde{%
\Lambda }$ of the hadronic molecule $\widetilde{\mathcal{M}}_{\mathrm{AV}}$
are carried out in accordance with a scheme described above. The QCD side of
relevant SRs are given by the expression
\begin{eqnarray}
&&\widetilde{\Pi }_{\mu \nu }^{\mathrm{OPE}}(p)=i\int d^{4}xe^{ipx}\left\{
\mathrm{Tr}\left[ \gamma _{5}S_{b}^{ab^{\prime }}(x)\gamma _{\nu
}S_{c}^{b^{\prime }b}(-x)\gamma _{\mu }\right. \right.  \notag \\
&&\left. \times S_{b}^{ba^{\prime }}(x)\gamma _{5}S_{b}^{a^{\prime }a}(-x)
\right] -\mathrm{Tr}\left[ \gamma _{5}S_{b}^{aa^{\prime }}(x)\gamma
_{5}S_{b}^{a^{\prime }a}(-x)\right]  \notag \\
&&\left. \times \mathrm{Tr}\left[ \gamma _{\mu }S_{b}^{bb^{\prime
}}(x)\gamma _{\nu }S_{c}^{b^{\prime }b}(-x)\right] \right\} ,
\end{eqnarray}%
whereas their phenomenological component can be obtained from Eq.\ (\ref%
{eq:Phys2}) by simple substitutions $m$, $\Lambda \rightarrow \widetilde{m}$%
, $\widetilde{\Lambda }$. Numerical computations confirm that SRs lead to
almost identical predictions for $\widetilde{m}$, $\widetilde{\Lambda }$:
Observed deviation, for instance, of $\ \widetilde{m}$ \ from $m$ amounts
approximately to $1-2\ \mathrm{MeV}$ which is beyond the accuracy of the
used SR method. Therefore, below we consider decays of the hadronic molecule
$\mathcal{M}_{\mathrm{AV}}$ and assume that $\widetilde{\mathcal{M}}_{%
\mathrm{AV}}$ has the same full width as $\mathcal{M}_{\mathrm{AV}}$.


\section{Decay of $\mathcal{M}_{\mathrm{AV}}$ to final states $\Upsilon
B_{c}^{-}$ and $\protect\eta _{b}B_{c}^{\ast -}$}

\label{sec:Widths1}

Here, we study dominant decays $\mathcal{M}_{\mathrm{AV}}\rightarrow
\Upsilon B_{c}^{-}$ and $\mathcal{M}_{\mathrm{AV}}\rightarrow \eta
_{b}B_{c}^{\ast -}$ of the hadronic molecule $\mathcal{M}_{\mathrm{AV}}$,
and calculate their partial widths. The result for the mass $m=(15800\pm 90)~%
\mathrm{MeV}$ of the molecule $\mathcal{M}_{\mathrm{AV}}$ shows that it
exceeds the two-meson thresholds for production of the pairs $\Upsilon
B_{c}^{-}$ and $\eta _{b}B_{c}^{\ast -}$. Indeed, the masses $m_{\Upsilon
}=(9460.40\pm 0.10)~\mathrm{MeV}$ and $m_{B_{c}}=(6274.47\pm 0.27\pm 0.17)~%
\mathrm{MeV}$ of the mesons $\Upsilon $ and $B_{c}^{-}$, as well as $m_{\eta
_{b}}=(9398.7\pm 2.0)~\mathrm{MeV}$ and $m_{B_{c}^{\ast }}=6338~\mathrm{MeV}$
of the mesons $\eta _{b}$ and $B_{c}^{\ast -}$ determine limits $15735~%
\mathrm{MeV}$ and $15737~\mathrm{MeV}$ which make possible dissociation to
these particles. Note that $m_{\Upsilon }$, $m_{B_{c}}$, and $m_{\eta _{b}}$
are experimental data \cite{PDG:2024}, whereas for $m_{B_{c}^{\ast }}$ we
use the model prediction \cite{Godfrey:2004ya}.

It is worth emphasizing that $\mathcal{M}_{\mathrm{AV}}$ decays through
these channels provided one employs for $m$ its central (or larger) value.
In the lower limit of the mass $m$ decays to pairs $\Upsilon B_{c}^{-}$ and $%
\eta _{b}B_{c}^{\ast -}$ are kinematically forbidden processes. Then to
estimate the full width of $\mathcal{M}_{\mathrm{AV}}$ one can consider only
subleading channels. This problem will be addressed later in this work, but
here we utilize as the mass of $\mathcal{M}_{\mathrm{AV}}$ the central value
in our prediction.


\subsection{Decay $\mathcal{M}_{\mathrm{AV}}\rightarrow \Upsilon B_{c}^{-}$}


The width of the process $\mathcal{M}_{\mathrm{AV}}\rightarrow \Upsilon
B_{c}^{-}$ contains, apart from other input parameters, also the strong
coupling $g_{1}$ at the vertex $\mathcal{M}_{\mathrm{AV}}\Upsilon B_{c}^{-}$%
. The coupling $g_{1}$ can be extracted at the mass shell $%
q^{2}=m_{B_{c}}^{2}$ of the form factor $g_{1}(q^{2})$. The latter is
evaluated using the SR obtained from analysis of the three-point correlator
\begin{eqnarray}
\Pi _{\mu \nu }^{1}(p,p^{\prime }) &=&i^{2}\int d^{4}xd^{4}ye^{ip^{\prime
}y}e^{-ipx}\langle 0|\mathcal{T}\{J_{\mu }^{\Upsilon }(y)  \notag \\
&&\times J^{B_{c}^{-}}(0)J_{\nu }^{\dagger }(x)\}|0\rangle ,  \label{eq:CF1a}
\end{eqnarray}%
where $J_{\mu }^{\Upsilon }(x)$ and $J^{B_{c}^{-}}(x)$ are the interpolating
currents of $\Upsilon $ and $B_{c}^{-}$, respectively. These currents are
given by the expressions
\begin{equation}
J_{\mu }^{\Upsilon }(x)=\overline{b}_{i}(x)\gamma _{\mu }b_{i}(x),\
J^{B_{c}^{-}}(x)=\overline{c}_{j}(x)i\gamma _{5}b_{j}(x).
\end{equation}%
The four-momentum $p$ of the molecule $\mathcal{M}_{\mathrm{AV}}$ is
connected to momenta $p^{\prime }$ and $q$ of produced particles, i.e., $%
p=p^{\prime }+q$.

The correlation function Eq.\ (\ref{eq:CF1a}) expressed using parameters of
the particles $\mathcal{M}_{\mathrm{AV}}$, $\Upsilon $ and $B_{c}^{-}$ is
the physical component $\Pi _{\mu \nu }^{1\mathrm{Phys}}(p,p^{\prime })$ of
the desired SR. To determine this correlator, we implement into Eq.\ (\ref%
{eq:CF1a}) systems of intermediate states for these particles and perform
four-dimensional integrations over $x$ and $y$. After separating the
contribution of the ground-level states, we get
\begin{eqnarray}
&&\Pi _{\mu \nu }^{1\mathrm{Phys}}(p,p^{\prime })=\frac{\langle 0|J_{\mu
}^{\Upsilon }|\Upsilon (p^{\prime },\varepsilon )\rangle }{p^{\prime
2}-m_{\Upsilon }^{2}}\frac{\langle 0|J^{B_{c}^{-}}|B_{c}^{-}(q)\rangle }{%
q^{2}-m_{B_{c}}^{2}}  \notag \\
&&\times \langle \Upsilon (p^{\prime },\varepsilon )B_{c}^{-}(q)|\mathcal{M}%
_{\mathrm{AV}}(p,\epsilon )\rangle \frac{\langle \mathcal{M}_{\mathrm{AV}%
}(p,\epsilon )|J_{\nu }^{\dagger }|0\rangle }{p^{2}-m^{2}}  \notag \\
&&+\cdots ,  \label{eq:TP1}
\end{eqnarray}%
with $\varepsilon _{\mu }$ being the polarization vector of $\Upsilon $. In
Eq.\ (\ref{eq:TP1}) the dots encode contributions of excited and continuum
states.

Having employed the matrix elements of the particles $\Upsilon $ and $%
B_{c}^{-}$%
\begin{eqnarray}
\langle 0|J_{\mu }^{\Upsilon }|\Upsilon (p^{\prime },\varepsilon )\rangle
&=&f_{\Upsilon }m_{\Upsilon }\varepsilon _{\mu },  \notag \\
\langle 0|J^{B_{c}^{-}}|B_{c}^{-}(q)\rangle &=&\frac{f_{B_{c}}m_{B_{c}}^{2}}{%
m_{b}+m_{c}},  \label{eq:ME1A}
\end{eqnarray}%
we rewrite $\Pi _{\mu \nu }^{1\mathrm{Phys}}(p,p^{\prime })$ in a form
suitable for our purposes. In Eq.\ (\ref{eq:ME1A}) $f_{\Upsilon }$ and $%
f_{B_{c}}$ are the decay constants of $\Upsilon $ and $B_{c}^{-}$,
respectively. We also employ an expression for the vertex $\langle \Upsilon
(p^{\prime },\varepsilon )B_{c}^{-}(q)|\mathcal{M}_{\mathrm{AV}}(p,\epsilon
)\rangle $
\begin{eqnarray}
&&\langle \Upsilon (p^{\prime },\varepsilon )B_{c}^{-}(q)|\mathcal{M}_{%
\mathrm{AV}}(p,\epsilon )\rangle =g_{1}(q^{2})\left[ (p\cdot q)(\epsilon
\cdot \varepsilon ^{\ast })\right.  \notag \\
&&\left. -(q\cdot \epsilon )(p\cdot \varepsilon ^{\ast })\right] .
\label{eq:AVVPS}
\end{eqnarray}%
By applying these matrix elements, it is not difficult to find that
\begin{eqnarray}
&&\Pi _{\mu \nu }^{1\mathrm{Phys}}(p,p^{\prime })=g_{1}(q^{2})\frac{\Lambda
f_{B_{c}}m_{B_{c}}^{2}f_{\Upsilon }m_{\Upsilon }}{(m_{b}+m_{c})\left(
p^{2}-m^{2}\right) \left( p^{\prime 2}-m_{\Upsilon }^{2}\right) }  \notag \\
&&\times \frac{1}{(q^{2}-m_{B_{c}}^{2})}\left[ \frac{m^{2}-m_{\Upsilon
}^{2}+q^{2}}{2}g_{\mu \nu }-p_{\mu }p_{\nu }+p_{\mu }^{\prime }p_{\nu
}\right.  \notag \\
&&\left. -\frac{m^{2}}{m_{\Upsilon }^{2}}p_{\mu }^{\prime }p_{\nu }^{\prime
}+\frac{m^{2}+m_{\Upsilon }^{2}-q^{2}}{m_{\Upsilon }^{2}}p_{\mu }^{\prime
}p_{\nu }\right] +\cdots .
\end{eqnarray}%
The function $\Pi _{\mu \nu }^{1\mathrm{Phys}}(p,p^{\prime })$ is a sum of
different Lorentz terms, one of which has be chosen for investigations. We
continue with the invariant amplitude $\Pi _{1}^{\mathrm{Phys}%
}(p^{2},p^{\prime 2},q^{2})$ that corresponds to the structure $g_{\mu \nu }$%
.

The correlation function $\Pi _{\mu \nu }^{1}(p,p^{\prime })$ in terms of
the quark propagators is equal to
\begin{eqnarray}
&&\Pi _{\mu \nu }^{1\mathrm{OPE}}(p,p^{\prime })=\int
d^{4}xd^{4}ye^{ip^{\prime }y}e^{-ipx}\left\{ \mathrm{Tr}\left[ \gamma _{\mu
}S_{b}^{ia}(y-x)\right. \right.  \notag \\
&&\left. \times \gamma _{\nu }S_{b}^{ai}(x-y)\right] \mathrm{Tr}\left[
\gamma _{5}S_{b}^{jb}(-x)\gamma _{5}S_{c}^{bj}(x)\right]  \notag \\
&&\left. -\mathrm{Tr}\left[ \gamma _{\mu }S_{b}^{ib}(y-x)\gamma
_{5}S_{c}^{bj}(x)\gamma _{5}S_{b}^{ja}(-x)\gamma _{\nu }S_{b}^{ai}(x-y)%
\right] \right\} .  \notag \\
&&  \label{eq:CF3}
\end{eqnarray}%
We label by $\Pi _{1}^{\mathrm{OPE}}(p^{2},p^{\prime 2},q^{2})$ the
invariant amplitude associated with the structure $\sim g_{\mu \nu }$ and
utilize it in our studies.

By equating $\Pi _{1}^{\mathrm{Phys}}(p^{2},p^{\prime 2},q^{2})$ and $\Pi
_{1}^{\mathrm{OPE}}(p^{2},p^{\prime 2},q^{2})$ and carrying out all required
technical manipulations (Borel transformations over the variables $-p^{2}$, $%
-p^{\prime 2}$, subtraction of excited and continuum states' contributions
), we determine SR for the form factor $g_{1}(q^{2})$
\begin{eqnarray}
&&g_{1}(q^{2})=\frac{2(m_{b}+m_{c})}{\Lambda
f_{B_{c}}m_{B_{c}}^{2}f_{\Upsilon }m_{\Upsilon }}\frac{q^{2}-m_{B_{c}}^{2}}{%
m^{2}-m_{\Upsilon }^{2}+q^{2}}  \notag \\
&&\times e^{m^{2}/M_{1}^{2}}e^{m_{\Upsilon }^{2}/M_{2}^{2}}\Pi _{1}(\mathbf{M%
}^{2},\mathbf{s}_{0},q^{2}),  \label{eq:SRCoupl1}
\end{eqnarray}%
where
\begin{eqnarray}
&&\Pi _{1}(\mathbf{M}^{2},\mathbf{s}_{0},q^{2})=%
\int_{(3m_{b}+m_{c})^{2}}^{s_{0}}ds\int_{4m_{b}^{2}}^{s_{0}^{\prime
}}ds^{\prime }\rho _{1}(s,s^{\prime },q^{2})  \notag \\
&&\times e^{-s/M_{1}^{2}}e^{-s^{\prime }/M_{2}^{2}}.  \label{eq:AS1}
\end{eqnarray}%
In Eqs.\ (\ref{eq:SRCoupl1}) and (\ref{eq:AS1}) $\Pi _{1}(\mathbf{M}^{2},%
\mathbf{s}_{0},q^{2})$ is the amplitude $\Pi _{1}^{\mathrm{OPE}%
}(p^{2},p^{\prime 2},q^{2})$ after the double Borel and continuum
subtraction procedures. The spectral density $\rho (s,s^{\prime },q^{2})$ in
the expression above is equal to the imaginary part of $\Pi _{1}^{\mathrm{OPE%
}}(s,s^{\prime },q^{2})$.

The function $\Pi _{1}(\mathbf{M}^{2},\mathbf{s}_{0},q^{2})$ contains
parameters $\mathbf{M}^{2}=(M_{1}^{2},M_{2}^{2})$ and $\mathbf{s}%
_{0}=(s_{0},s_{0}^{\prime })$ where $(M_{1}^{2},s_{0})$ and $%
(M_{2}^{2},s_{0}^{\prime })$ correspond to channels of the particles $%
\mathcal{M}_{\mathrm{AV}}$ and $\Upsilon $. They should be constrained in
accordance with usual rules of SR computations which have been explained in
the previous section. Numerical analysis proves that Eq.\ (\ref{eq:Wind1})
for the parameters $(M_{1}^{2},s_{0})$ and
\begin{equation}
M_{2}^{2}\in \lbrack 10,12]~\mathrm{GeV}^{2},\ s_{0}^{\prime }\in \lbrack
98,100]~\mathrm{GeV}^{2}.  \label{eq:Wind3}
\end{equation}%
for $(M_{2}^{2},s_{0}^{\prime })$ satisfy these constraints.

\begin{figure}[h]
\includegraphics[width=8.5cm]{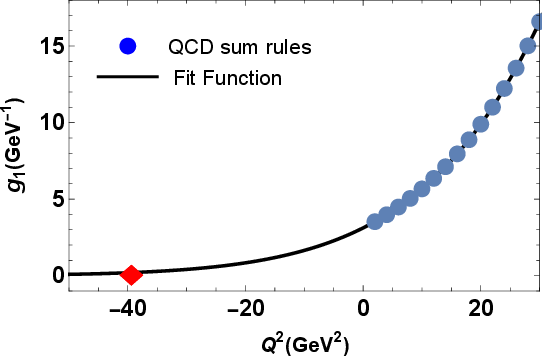}
\caption{QCD data and extrapolating function $\mathcal{F}_1 (Q^{2})$. The
diamond fixes the point $Q^{2}=-m_{B_c}^{2}$. }
\label{fig:Fit}
\end{figure}

For numerical computations one needs also the spectroscopic parameters of
the mesons $\Upsilon $ and $B_{c}^{-}$. There are experimental information
on masses of these particles \cite{PDG:2024}. Their decay constants $%
f_{\Upsilon }=(708\pm 8)~\mathrm{MeV}$ and $\ f_{B_{c}}=(371\pm 37)~\mathrm{%
MeV}$ are borrowed from Refs.\ \cite{Lakhina:2006vg,Wang:2024fwc},
respectively. The SR approach generates reliable predictions in the
Euclidean region $q^{2}<0$, whereas $g_{1}(q^{2})$ becomes equal to $g_{1}$
at the mass shell $q^{2}=m_{B_{c}}^{2}$. Therefore, it is convenient to
introduce the new function $g_{1}(Q^{2})$, where $Q^{2}=-q^{2}$, and apply
it in future studies. The QCD predictions for $g_{1}(Q^{2})$ are depicted in
Fig.\ \ref{fig:Fit}, where $Q^{2}$ changes inside limits $Q^{2}=2-30~\mathrm{%
GeV}^{2}$.

To extract $g_{1}$ at $q^{2}=-Q^{2}=m_{B_{c}}^{2}$, we introduce the fit
function $\mathcal{F}_{1}(Q^{2})$: For $Q^{2}>0$ it amounts to the QCD data,
but can also be extended to the region of $Q^{2}<0$. We choose this function
in the following form
\begin{equation}
\mathcal{F}_{i}(Q^{2})=\mathcal{F}_{i}^{0}\mathrm{\exp }\left[ l_{i}^{1}%
\frac{Q^{2}}{m^{2}}+l_{i}^{2}\left( \frac{Q^{2}}{m^{2}}\right) ^{2}\right] ,
\label{eq:FitF}
\end{equation}%
where the parameters $\mathcal{F}_{i}^{0}$, $l_{i}^{1}$, and $l_{i}^{2}$
have to be extracted from comparison of $\mathcal{F}_{1}(Q^{2})$ and the QCD
data. It is easy to find them:
\begin{equation}
\mathcal{F}_{1}^{0}=3.21~\mathrm{GeV}^{-1},l_{1}^{1}=15.40,\text{and }%
l_{1}^{2}=-12.40.  \label{eq:FF1}
\end{equation}%
In Fig.\ \ref{fig:Fit} we also show $\mathcal{F}_{1}(Q^{2})$: A reasonable
agreement of $\mathcal{F}_{1}(Q^{2})$ and QCD data is clear. As a result,
for $g_{1}$ we find
\begin{equation}
g_{1}\equiv \mathcal{F}_{1}(-m_{B_{c}}^{2})=(2.0\pm 0.4)\times 10^{-1}\
\mathrm{GeV}^{-1}.  \label{eq:g1}
\end{equation}

The width of the process $\mathcal{M}_{\mathrm{AV}}\rightarrow \Upsilon
B_{c}^{-}$ can be calculated by means of the expression
\begin{equation}
\Gamma \left[ \mathcal{M}_{\mathrm{AV}}\rightarrow \Upsilon B_{c}^{-}\right]
=g_{1}^{2}\frac{\lambda _{1}}{24\pi m^{2}}|M_{1}|^{2},  \label{eq:PDw1}
\end{equation}%
where
\begin{eqnarray}
|M_{1}|^{2} &=&\frac{1}{4m_{\Upsilon }^{2}}\left[
m^{6}-2m^{4}m_{B_{c}}^{2}+2m_{\Upsilon }^{2}(m_{\Upsilon
}^{2}-m_{B_{c}}^{2})^{2}\right.  \notag \\
&&\left. +m^{2}(m_{B_{c}}^{4}+6m_{B_{c}}^{2}m_{\Upsilon }^{2}-3m_{\Upsilon
}^{4})\right] .
\end{eqnarray}%
We have used also the parameter $\lambda _{1}=\lambda (m,m_{\Upsilon
},m_{B_{c}})$ with $\lambda (a,b,c)$ being defined as
\begin{equation}
\lambda (a,b,c)=\frac{\sqrt{%
a^{4}+b^{4}+c^{4}-2(a^{2}b^{2}+a^{2}c^{2}+b^{2}c^{2})}}{2a}.
\end{equation}

Finally, we get,
\begin{equation}
\Gamma \left[ \mathcal{M}_{\mathrm{AV}}\rightarrow \Upsilon B_{c}^{-}\right]
=(46.9\pm 13.3)~\mathrm{MeV}.  \label{eq:DW2}
\end{equation}

\subsection{Process $\mathcal{M}_{\mathrm{AV}}\rightarrow \protect\eta %
_{b}B_{c}^{\ast -}$}


The process $\mathcal{M}_{\mathrm{AV}}\rightarrow \eta _{b}B_{c}^{\ast -}$
is studied in accordance with the scheme outlined above. Here, we are going
to estimate the coupling $g_{2}$ at $\mathcal{M}_{\mathrm{AV}}\eta
_{b}B_{c}^{\ast -}$. For these purposes, we start from analysis of the
correlation function $\Pi _{\mu \nu }^{2}(p,p^{\prime })$ that should allow
us to evaluate the form factor $g_{2}(q^{2})$: The latter at the mass shell $%
q^{2}=m_{B_{c}^{\ast }}^{2}$ gives $g_{2}$.

This correlation function is introduced by means of the formula
\begin{eqnarray}
\Pi _{\mu \nu }^{2}(p,p^{\prime }) &=&i^{2}\int d^{4}xd^{4}ye^{ip^{\prime
}y}e^{-ipx}\langle 0|\mathcal{T}\{J^{\eta _{b}}(y)  \notag \\
&&\times J_{\mu }^{B_{c}^{\ast }}(0)J_{\nu }^{\dagger }(x)\}|0\rangle ,
\end{eqnarray}%
with $J^{\eta _{b}}(x)$ and $J_{\mu }^{B_{c}^{\ast }}(x)$ being the currents
that interpolate particles $\eta _{b}$ and $B_{c}^{\ast -}$. They have the
following forms
\begin{equation}
J^{\eta _{b}}(x)=\overline{b}_{i}(x)i\gamma _{5}b_{i}(x),\ J_{\mu
}^{B_{c}^{\ast }}(x)=\overline{c}_{j}(x)\gamma _{\mu }b_{j}(x).
\end{equation}%
The correlator $\Pi _{\mu \nu }^{2}(p,p^{\prime })$ in terms of physical
parameters of the particles $\mathcal{M}_{\mathrm{AV}}$, $\eta _{b}$, and $%
B_{c}^{\ast -}$ is given by the expression%
\begin{eqnarray}
&&\Pi _{\mu \nu }^{2\mathrm{Phys}}(p,p^{\prime })=\frac{g_{2}(q^{2})\Lambda
f_{\eta _{b}}m_{\eta _{b}}^{2}f_{B_{c}^{\ast }}m_{B_{c}^{\ast }}}{%
2m_{b}\left( p^{2}-m^{2}\right) \left( p^{\prime 2}-m_{\eta _{b}}^{2}\right)
(q^{2}-m_{B_{c}^{\ast }}^{2})}  \notag \\
&&\times \left[ \frac{m^{2}+m_{\eta _{b}}^{2}-q^{2}}{2}g_{\mu \nu }-\frac{%
m^{2}}{m_{B_{c}^{\ast }}^{2}}p_{\mu }^{\prime }p_{\nu }^{\prime }+\frac{%
m^{2}-m_{B_{c}^{\ast }}^{2}}{m_{B_{c}^{\ast }}^{2}}p_{\mu }^{\prime }p_{\nu
}\right.  \notag \\
&&\left. -\frac{m^{2}+m_{\eta _{b}}^{2}-q^{2}}{2m_{B_{c}^{\ast }}^{2}}%
(p_{\mu }p_{\nu }+p_{\mu }p_{\nu }^{\prime })\right] +\cdots .
\end{eqnarray}%
To derive $\Pi _{\mu \nu }^{2\mathrm{Phys}}(p,p^{\prime })$ we have employed
the matrix elements
\begin{eqnarray}
\langle 0|J_{\mu }^{B_{c}^{\ast }}|B_{c}^{\ast -}(q,\varepsilon )\rangle
&=&f_{B_{c}^{\ast }}m_{B_{c}^{\ast }}\varepsilon _{\mu },  \notag \\
\langle 0|J^{\eta _{b}}|\eta _{b}(p^{\prime })\rangle &=&\frac{f_{\eta
_{b}}m_{\eta _{b}}^{2}}{2m_{b}},
\end{eqnarray}%
and
\begin{eqnarray}
&&\langle \eta _{b}(p^{\prime })B_{c}^{\ast -}(q,\varepsilon )|\mathcal{M}_{%
\mathrm{AV}}(p,\epsilon )\rangle =g_{2}(q^{2})\left[ (p\cdot p^{\prime
})(\epsilon \cdot \varepsilon ^{\ast })\right.  \notag \\
&&\left. -(p^{\prime }\cdot \epsilon )(p\cdot \varepsilon ^{\ast }\right] .
\end{eqnarray}%
The QCD side of the SR is determined by the formula
\begin{eqnarray}
&&\Pi _{\mu \nu }^{2\mathrm{OPE}}(p,p^{\prime })=-\int
d^{4}xd^{4}ye^{ip^{\prime }y}e^{-ipx}\mathrm{Tr}\left[ \gamma
_{5}S_{b}^{ib}(y-x)\right.  \notag \\
&&\left. \times \gamma _{5}S_{c}^{bj}(x)\gamma _{\mu }S_{b}^{ja}(-x)\gamma
_{\nu }S_{b}^{ai}(x-y)\right] .
\end{eqnarray}%
To find SR for the form factor $g_{2}(q^{2})$ we use the invariant
amplitudes $\Pi _{2}^{\mathrm{Phys}}(p^{2},p^{\prime 2},q^{2})$ and $\Pi
_{2}^{\mathrm{OPE}}(p^{2},p^{\prime 2},q^{2})$ related to structures $\sim
g_{\mu \nu }$ both in the phenomenological and $\mathrm{OPE}$ versions of
the correlation function $\Pi _{\mu \nu }^{2}(p,p^{\prime })$. Then, the
required SR reads
\begin{eqnarray}
&&g_{2}(q^{2})=\frac{4m_{b}}{\Lambda f_{\eta _{b}}m_{\eta
_{b}}^{2}f_{B_{c}^{\ast }}m_{B_{c}^{\ast }}}\frac{q^{2}-m_{B_{c}^{\ast }}^{2}%
}{m^{2}+m_{\eta _{b}}^{2}-q^{2}}  \notag \\
&&\times e^{m^{2}/M_{1}^{2}}e^{m_{\eta _{b}}^{2}/M_{2}^{2}}\Pi _{2}(\mathbf{M%
}^{2},\mathbf{s}_{0},q^{2}).
\end{eqnarray}

\begin{figure}[h]
\includegraphics[width=8.5cm]{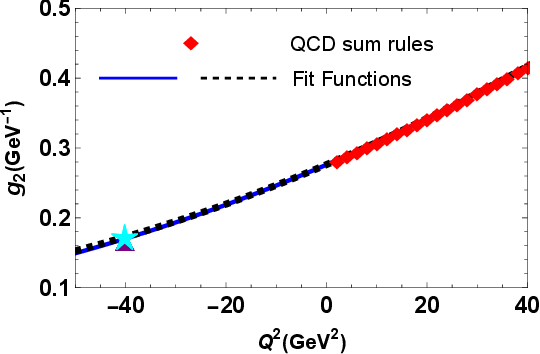}
\caption{SR results and fitting functions $\mathcal{F}_2 (Q^{2})$ (solid
line) and $\mathcal{F}_{2A} (Q^{2})$ (dashed line). The star and triangle
are at $Q^{2}=-m_{B_{c}^{\ast}}^2$. }
\label{fig:FitA}
\end{figure}

Numerical computations are carried out by employing the decay constants $%
f_{\eta _{b}}=724~\mathrm{MeV}$ and $f_{B_{c}^{\ast }}=471~\mathrm{MeV}$
\cite{Eichten:2019gig}, respectively. The regions
\begin{equation}
M_{2}^{2}\in \lbrack 10,12]~\mathrm{GeV}^{2},\ s_{0}^{\prime }\in \lbrack
95,99]~\mathrm{GeV}^{2},
\end{equation}%
in the channel of $\eta _{b}$ meson meet necessary restrictions of SR
analysis. The relevant extrapolating function $\mathcal{F}_{2}(Q^{2})$ has
the parameters
\begin{equation}
\mathcal{F}_{2}^{0}=0.276~\mathrm{GeV}^{-1},l_{2}^{1}=2.793,\text{and }%
l_{2}^{2}=-1.384.
\end{equation}%
The SR data and the function $\mathcal{F}_{2}(Q^{2})$ are plotted in Fig.\ %
\ref{fig:FitA}. We obtain for $g_{2}$%
\begin{equation}
g_{2}\equiv \mathcal{F}_{2}(-m_{B_{c}}^{2})=(1.69\pm 0.30)\times 10^{-1}\
\mathrm{GeV}^{-1}.  \label{eq:g2}
\end{equation}

To evaluate the width of this decay we employ the following expression
\begin{equation}
\Gamma \left[ \mathcal{M}_{\mathrm{AV}}\rightarrow \eta _{b}B_{c}^{\ast -}%
\right] =g_{2}^{2}\frac{\lambda _{2}}{24\pi m^{2}}|M_{2}|^{2},
\end{equation}%
where $|M_{2}|$ is obtainable from $|M_{1}|$ upon $m_{B_{c}}$, $m_{\Upsilon
}\rightarrow m_{\eta _{b}}$, $m_{B_{c}^{\ast }}$ substitutions. Now $\lambda
_{2}$ amounts to $\lambda (m,m_{\eta _{b}},m_{B_{c}^{\ast }})$. The width of
this process is
\begin{equation}
\Gamma \left[ \mathcal{M}_{\mathrm{AV}}\rightarrow \eta _{b}B_{c}^{\ast -}%
\right] =(33.3\pm 9.5)~\mathrm{MeV}.
\end{equation}

While considering the decay channels $\mathcal{M}_{\mathrm{AV}}\rightarrow
\Upsilon B_{c}^{-}$ and $\eta _{b}B_{c}^{\ast -}$ we have applied the
function Eq.\ (\ref{eq:FitF}) to extrapolate SR data to region of $Q^{2}<0$
which is necessary for estimation of the strong couplings at the relevant
three-particle vertices. But this procedure can be fulfilled using
alternative fit functions: A choice of a new function, in general, may
modify results. To explore this problem, we introduce the function
\begin{equation}
\mathcal{F}_{2A}(Q^{2})=\frac{f_{0}\left( 1-Q^{2}/m^{2}\right) ^{-2}}{\left[
1-s_{1}(Q^{2}/m^{2})+s_{2}\left( Q^{4}/m^{4}\right) \right] },
\label{eq:FitF2}
\end{equation}%
and compute the strong coupling $g_{2}$. Here, $f_{0}$, $s_{1}$ and $s_{2}$
are fitting constants. By employing SR data and Eq. (\ref{eq:FitF2}) one
gets $f_{0}=0.276~\mathrm{GeV}^{-1}$, $s_{1}=0.775$ and $s_{2}=2.486$. In
Fig.\ \ref{fig:FitA}\ we demonstrate $\mathcal{F}_{2A}(Q^{2})$, where its
nice agreement with SQ data, as well as with $\mathcal{F}_{2}(Q^{2})$ is
evident. The extrapolating function $\mathcal{F}_{2A}(Q^{2})$ leads to the
following prediction for the strong coupling $g_{2}=0.172~\mathrm{GeV}^{-1}$%
. The deviation $|0.003|$ of this value from one presented in Eq.\ (\ref%
{eq:g2}) is an order of magnitude smaller that uncertainties $\pm 0.03$ of $%
g_{2}$. Therefore, throughout this article we utilize Eq.\ (\ref{eq:FitF})
and neglect small ambiguities connected with a choice of other extrapolating
functions.

\section{Decays generated by annihilation of $\overline{b}b $ quarks}

\label{sec:Widths2}


Annihilation of $\overline{b}b$ quarks triggers various decay modes of the
hadronic molecular state $\mathcal{M}_{\mathrm{AV}}$. Decays $\mathcal{M}_{%
\mathrm{AV}}\rightarrow B^{\ast -}\overline{D}^{0}$, $\overline{B}^{\ast
0}D^{-}$, $B^{-}\overline{D}^{\ast 0}$, $\overline{B}^{0}D^{\ast -}$, $%
\overline{B}_{s}^{\ast 0}D_{s}^{-}$, and $\overline{B}_{s}^{0}D_{s}^{\ast -}$
are such channels. Note that parameters some of these processes are very
close to each other. This is connected with the following fact: In this work
we use the approximation $m_{\mathrm{u}}=m_{\mathrm{d}}=0$ and $m_{\mathrm{s}%
}=(93.5\pm 0.8)~\mathrm{MeV}$. Then it is not difficult to see that the QCD
sides of the SRs for the channels $\mathcal{M}_{\mathrm{AV}}\rightarrow
B^{\ast -}\overline{D}^{0}$ and $\overline{B}^{\ast 0}D^{-}$, as well as for
$\mathcal{M}_{\mathrm{AV}}\rightarrow B^{-}\overline{D}^{\ast 0}$and $%
\overline{B}^{0}D^{\ast -}$ are given by the same expressions. Differences
in the physical sides are connected with the masses of mesons, for example, $%
\overline{D}^{0}$ and $D^{-}$ which are small. Therefore, we consider these
processes as identical channels including them into two groups and calculate
the partial width one of them.


\subsection{$\mathcal{M}_{\mathrm{AV}}\rightarrow B^{\ast -}\overline{D}^{0}$%
, $\overline{B}^{\ast 0}D^{-}$}


In this subsection, we investigate the channel $\mathcal{M}_{\mathrm{AV}%
}\rightarrow B^{\ast -}\overline{D}^{0}$ of the molecule $\mathcal{M}_{%
\mathrm{AV}}$ and compute its partial decay width. We are going to estimate
the coupling $G$ at the vertex $\mathcal{M}_{\mathrm{AV}}B^{\ast -}\overline{%
D}^{0}$. The correlation function suitable for our purposes is
\begin{eqnarray}
\overline{\Pi }_{\mu \nu }(p,p^{\prime }) &=&i^{2}\int
d^{4}xd^{4}ye^{ip^{\prime }y}e^{-ipx}\langle 0|\mathcal{T}\{J_{\mu
}^{B^{\ast }}(y)  \notag \\
&&\times J^{\overline{D}^{0}}(0)J_{\nu }^{\dagger }(x)\}|0\rangle ,
\label{eq:CF3A}
\end{eqnarray}%
where $J_{\mu }^{B^{\ast }}$ and $J^{\overline{D}^{0}}(x)$ are currents for $%
B^{\ast -}$ and $\overline{D}^{0}$ mesons which are given by the expressions
\begin{equation}
J_{\mu }^{B^{\ast }}(x)=\overline{u}_{i}(x)\gamma _{\mu }b_{i}(x),\ J^{%
\overline{D}^{0}}(x)=\overline{c}_{j}(x)i\gamma _{5}u_{j}(x).
\end{equation}

The phenomenological expression for $\overline{\Pi }_{\mu \nu }(p,p^{\prime
})$ is obtained using the matrix elements
\begin{eqnarray}
\langle 0|J_{\mu }^{B^{\ast }}|B^{\ast -}(p^{\prime },\varepsilon )\rangle
&=&f_{B^{\ast }}m_{B^{\ast }}\varepsilon _{\mu },  \notag \\
\langle 0|J^{\overline{D}^{0}}|\overline{D}^{0}(q)\rangle &=&\frac{f_{D}m_{%
\overline{D}^{0}}^{2}}{m_{c}}.  \label{eq:ME3}
\end{eqnarray}%
In Eq.\ (\ref{eq:ME3}) $m_{B^{\ast }}=(5324.75\pm 0.20)~\mathrm{MeV},\ m_{%
\overline{D}^{0}}=(1864.84\pm 0.05)~\mathrm{MeV}$ and $f_{B^{\ast }}=(210\pm
6)~\mathrm{MeV},$ $f_{D}=(211.9\pm 1.1)~\mathrm{MeV}$ are the masses and
decay constants of  these mesons, whereas $\varepsilon _{\mu }$ is the
polarization vector of the vector meson $B^{\ast -}$. The vertex $\langle B^{\ast -}(p^{\prime },\varepsilon )\overline{D}^{0}(q)|\mathcal{M}_{
\mathrm{AV}}(p,\epsilon )\rangle $ and correlator $\overline{\Pi }_{\mu \nu }^{\mathrm{%
Phys}}(p,p^{\prime })$ are analogous to ones presented in section \ref%
{sec:Widths1}.

Calculations of $\overline{\Pi }_{\mu \nu }(p,p^{\prime })$ using the quark
propagators yield
\begin{eqnarray}
&&\overline{\Pi }_{\mu \nu }^{\mathrm{OPE}}(p,p^{\prime })=\frac{\langle
\overline{b}b\rangle }{3}\int d^{4}xd^{4}ye^{ip^{\prime }y}e^{-ipx}\mathrm{Tr%
}\left[ \gamma _{\mu }S_{b}^{ia}(y-x)\right.  \notag \\
&&\left. \times \gamma _{\nu }\gamma _{5}S_{c}^{aj}(x)\gamma
_{5}S_{u}^{ji}(-y)\right] .
\end{eqnarray}%
We use also the relation
\begin{equation}
\langle \overline{b}b\rangle =-\frac{1}{12m_{b}}\langle \frac{\alpha
_{s}G^{2}}{\pi }\rangle  \label{eq:Conden}
\end{equation}%
obtained in Ref.\ \cite{Shifman:1978bx} in the framework of SR method.

\begin{figure}[h]
\includegraphics[width=8.5cm]{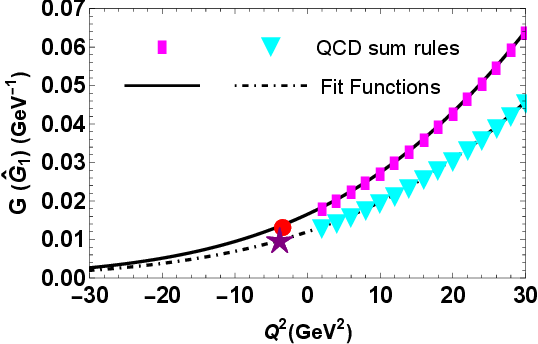}
\caption{SR data for the form factors $G(Q^{2})$ and $\widehat{G}_{1}(Q^2)$
and fit functions $\overline{\mathcal{F}}(Q^{2})$ (solid line), $\widehat{%
\mathcal{F}}_1(Q^{2})$ (dash-dotted line). The circle and star are placed at
positions $Q^{2}=-m_{\overline{D}^{0}}^{2}$ and $Q^{2}=-m_{D_{s}^{-}}^{2}$,
respectively. }
\label{fig:Fit1}
\end{figure}

The form factor $G(Q^{2})$ that at $Q^{2}=-m_{\overline{D}^{0}}^{2}$ amounts
to $G$ is computed in the region $Q^{2}=2-30\ \mathrm{GeV}^{2}$. For $%
(M_{1}^{2},s_{0})$ we have used their values from Eq.\ (\ref{eq:Wind1}), and
the parameters $(M_{2}^{2},s_{0}^{\prime })$ have been varied within the
borders
\begin{equation}
M_{2}^{2}\in \lbrack 5.5,6.5]~\mathrm{GeV}^{2},\ s_{0}^{\prime }\in \lbrack
34,35]~\mathrm{GeV}^{2}.
\end{equation}%
Results for $G(Q^{2})$ are plotted in Fig.\ \ref{fig:Fit1}. The
extrapolating function $\overline{\mathcal{F}}(Q^{2})$ is determined by the
constants $\overline{\mathcal{F}}^{0}=0.0166~\mathrm{GeV}^{-1}$, $\overline{l%
}^{1}=13.28$, and $\overline{l}^{2}=-16.70.$ The strong coupling $G$ is
evaluated at the mass shell $q^{2}=m_{\overline{D}^{0}}^{2}$ and equals to%
\begin{equation}
G\equiv \overline{\mathcal{F}}(-m_{\overline{D}^{0}}^{2})=(1.37\pm
0.31)\times 10^{-2}~\mathrm{GeV}^{-1}.
\end{equation}%
For the width of the process $\mathcal{M}_{\mathrm{AV}}\rightarrow B^{\ast -}%
\overline{D}^{0}$ our analysis predicts
\begin{equation}
\Gamma \left[ \mathcal{M}_{\mathrm{AV}}\rightarrow B^{\ast -}\overline{D}^{0}%
\right] =(8.9\pm 2.9)~\mathrm{MeV}.
\end{equation}%
Let us note that ambiguities above are total errors generated by
uncertainties in $G$ and in the masses of the particles $\mathcal{M}_{%
\mathrm{AV}}$, $B^{\ast -}$, and $\overline{D}^{0}$.

The width of the second process $\mathcal{M}_{\mathrm{AV}}\rightarrow
\overline{B}^{\ast 0}D^{-}$, as it has been discussed above, is
approximately equal to that of the first decay. In other words%
\begin{equation}
\Gamma \left[ \mathcal{M}_{\mathrm{AV}}\rightarrow \overline{B}^{\ast 0}D^{-}%
\right] \approx \Gamma \left[ \mathcal{M}_{\mathrm{AV}}\rightarrow B^{\ast -}%
\overline{D}^{0}\right] .
\end{equation}

\subsection{$\mathcal{M}_{\mathrm{AV}}\rightarrow B^{-}\overline{D}^{\ast 0}$%
, $\overline{B}^{0}D^{\ast -}$}


These channels are studied within the same approach. Here, we give formulas
for the decay $\mathcal{M}_{\mathrm{AV}}\rightarrow B^{-}\overline{D}^{\ast
0}$ and our estimate for its partial width: The partial width of the second mode
$\mathcal{M}_{\mathrm{AV}}\rightarrow \overline{B}^{0}D^{\ast -}$ is
approximately equal to parameters of the first decay.

We start from studying the correlation function
\begin{eqnarray}
\widetilde{\Pi }_{\mu \nu }(p,p^{\prime }) &=&i^{2}\int
d^{4}xd^{4}ye^{ip^{\prime }y}e^{-ipx}\langle 0|\mathcal{T}\{J^{B}(y)  \notag
\\
&&\times J_{\mu }^{\overline{D}^{\ast 0}}(0)J_{\nu }^{\dagger
}(x)\}|0\rangle .  \label{eq:CF2}
\end{eqnarray}%
Here, $J^{B}(x)$ and $J_{\mu }^{\overline{D}^{\ast 0}}(x)$ are relevant
interpolating currents%
\begin{equation}
J^{B}(x)=\overline{u}_{i}(x)i\gamma _{5}b_{i}(x),\ J_{\mu }^{\overline{D}%
^{\ast 0}}(x)=\overline{c}_{j}(x)\gamma _{\mu }u_{j}(x).
\end{equation}%
The correlator in Eq.\ (\ref{eq:CF2}) is required to find SR for the form
factor $\widetilde{G}(q^{2})$ that at the mass shell of the meson $\overline{%
D}^{\ast 0}$ permits us to evaluate the strong coupling $\widetilde{G}$ at
the vertex $\mathcal{M}_{\mathrm{AV}}\overline{B}^{0}D^{\ast -}$.

We calculate the correlator $\widetilde{\Pi }_{\mu \nu }^{\mathrm{Phys}%
}(p,p^{\prime })$ by employing the following matrix elements
\begin{eqnarray}
\langle 0|J^{B}|B^{-}(p^{\prime })\rangle &=&\frac{f_{B}m_{B}^{2}}{m_{b}},
\notag \\
\langle 0|J_{\mu }^{\overline{D}^{\ast 0}}|\overline{D}^{\ast
0}(q,\varepsilon )\rangle &=&f_{\overline{D}^{\ast 0}}m_{\overline{D}^{\ast
0}}\varepsilon _{\mu },  \label{eq:ME4}
\end{eqnarray}%
where $m_{B}=(5279.41\pm 0.07)~\mathrm{MeV}$, $\ f_{B}=206~\mathrm{MeV}$ and
$m_{\overline{D}^{\ast 0}}=(2006.85\pm 0.05)~\mathrm{MeV}$, $f_{\overline{D}%
^{\ast 0}}=(252.2\pm 22.66)~\mathrm{MeV}$ are spectroscopic parameters of
the mesons under discussion. The polarization vector of the $\overline{D}%
^{\ast 0}$ meson is labeled by $\varepsilon _{\mu }$.

The QCD component of SR is found in the form
\begin{eqnarray}
&&\widetilde{\Pi }_{\mu \nu }^{\mathrm{OPE}}(p,p^{\prime })=\frac{\langle
\overline{b}b\rangle }{3}\int d^{4}xd^{4}ye^{ip^{\prime }y}e^{-ipx}\mathrm{Tr%
}\left[ \gamma _{5}S_{b}^{ia}(y-x)\right.  \notag \\
&&\left. \times \gamma _{\nu }\gamma _{5}S_{c}^{aj}(x)\gamma _{\mu
}S_{u}^{ji}(-y)\right] .
\end{eqnarray}%
Numerical computations of SR for $\widetilde{G}(Q^{2})$ fulfilled at $%
Q^{2}=2-30\ \mathrm{GeV}^{2}$ using the regions Eq.\ (\ref{eq:Wind1}) for $%
(M_{1}^{2},s_{0})$ and
\begin{equation}
M_{2}^{2}\in \lbrack 5.5,6.5]~\mathrm{GeV}^{2},\ s_{0}^{\prime }\in \lbrack
33.5,34.5]~\mathrm{GeV}^{2},
\end{equation}%
for the parameters $(M_{2}^{2},s_{0}^{\prime })$. The fit function $%
\widetilde{\mathcal{F}}(Q^{2})$ that employed to fix the coupling $%
\widetilde{G}$ has the following parameters $\widetilde{\mathcal{F}}%
^{0}=0.0049~\mathrm{GeV}^{-1},\widetilde{l}^{1}=12.96,\widetilde{l}%
^{2}=-16.96.$ This function leads to the predictions

\begin{equation}
\widetilde{G}\equiv \widetilde{\mathcal{F}}(-m_{\overline{D}^{\ast
0}}^{2})=(3.94\pm 0.78)\times 10^{-3}\ \mathrm{GeV}^{-1},
\end{equation}%
and
\begin{equation}
\Gamma \left[ \mathcal{M}_{\mathrm{AV}}\rightarrow B^{-}\overline{D}^{\ast 0}%
\right] =(4.4\pm 1.3)~\mathrm{MeV}.  \label{eq:DW4}
\end{equation}%
The partial width of the process $\mathcal{M}_{\mathrm{AV}}\rightarrow
\overline{B}^{0}D^{\ast -}$ also amounts to Eq.\ (\ref{eq:DW4}).


\subsection{$\mathcal{M}_{\mathrm{AV}}\rightarrow \overline{B}_{s}^{\ast
0}D_{s}^{-}$, $\overline{B}_{s}^{0}D_{s}^{\ast -}$}


Remaining two modes of $\mathcal{M}_{\mathrm{AV}}$ for investigations are
its decays to the pairs of mesons $\overline{B}_{s}^{\ast 0}D_{s}^{-}$ and $%
\overline{B}_{s}^{0}D_{s}^{\ast -}$. We consider in detail the channel $%
\mathcal{M}_{\mathrm{AV}}\rightarrow \overline{B}_{s}^{\ast 0}D_{s}^{-}$ and
provide results for the second process.

For the decay $\mathcal{M}_{\mathrm{AV}}\rightarrow \overline{B}_{s}^{\ast
0}D_{s}^{-}$ one should analyze the correlation function
\begin{eqnarray}
\widehat{\Pi }_{\mu \nu }^{1}(p,p^{\prime }) &=&i^{2}\int
d^{4}xd^{4}ye^{ip^{\prime }y}e^{-ipx}\langle 0|\mathcal{T}\{J_{\mu
}^{B_{s}^{\ast }}(y)  \notag \\
&&\times J^{D_{s}}(0)J_{\nu }^{\dagger }(x)\}|0\rangle ,  \label{eq:CF3B}
\end{eqnarray}%
which allows us to get SR for the form factor $\widehat{G}_{1}(Q^{2})$. In
Eq.\ (\ref{eq:CF3B})$\ J_{\mu }^{B_{s}^{\ast }}(x)$ and $J^{D_{s}}(x)$ are
interpolating currents for the mesons $\overline{B}_{s}^{\ast 0}$ and $%
D_{s}^{-}$%
\begin{equation}
J_{\mu }^{B_{s}^{\ast }}(x)=\overline{s}_{i}(x)\gamma _{\mu
}b_{i}(x),~J^{D_{s}}(x)=\overline{c}_{j}(x)i\gamma _{5}s_{j}(x).
\end{equation}

To find the phenomenological component of SR we apply the matrix elements%
\begin{eqnarray}
&&\langle 0|J_{\mu }^{B^{\ast }_s}|\overline{B}_{s}^{\ast 0}(p^{\prime
},\varepsilon )\rangle =f_{B_{s}^{\ast }}m_{B_{s}^{\ast }}\varepsilon _{\mu
},  \notag \\
&&\langle 0|J^{D_{s}}|D_{s}^{-}(q)\rangle =\frac{f_{D_{s}}m_{D_{s}}^{2}}{%
m_{c}+m_{s}}.  \label{eq:ME5}
\end{eqnarray}%
The vertex $\langle \overline{B}_{s}^{\ast 0}(p^{\prime
},\varepsilon )D_{s}^{-}(q)|\mathcal{M}_{\mathrm{AV}}(p,\epsilon )\rangle  $ has the
standard form. In Eq.\ (\ref{eq:ME5}) $m_{B_{s}^{\ast }}=(5415.4\pm 1.4)~%
\mathrm{MeV}$, $f_{B_{s}^{\ast }}=221~\mathrm{MeV}$ and $m_{D_{s}}=(1968.35%
\pm 0.07)~\mathrm{MeV}$, $f_{D_{s}}=(249.9\pm 0.5)~\mathrm{MeV}$ are the
parameters of the final-state mesons.

The correlator $\widehat{\Pi }_{\mu \nu }^{1}(p,p^{\prime })$ in terms of
the heavy quark propagators is
\begin{eqnarray}
\widehat{\Pi }_{\mu \nu }^{1\mathrm{OPE}}(p,p^{\prime }) &=&\frac{\langle
\overline{b}b\rangle }{3}\int d^{4}xd^{4}ye^{ip^{\prime }y}e^{-ipx}\mathrm{Tr%
}\left[ \gamma _{\mu }S_{b}^{ia}(y-x)\right.  \notag \\
&&\left. \times \gamma _{\nu }\gamma _{5}S_{c}^{aj}(x)\gamma
_{5}S_{s}^{ji}(-y)\right] .
\end{eqnarray}%
The SR for the form factor $\widehat{G}_{1}(Q^{2})$ is found by utilizing
amplitudes connected with the terms $\sim g_{\mu \nu }$ both in $\widehat{\Pi} _{\mu
\nu }^{\mathrm{1Phys}}(p,p^{\prime })$ and $\widehat{\Pi} _{\mu \nu }^{\mathrm{1OPE}%
}(p,p^{\prime })$.

Numerical calculations are performed by employing
\begin{equation}
M_{2}^{2}\in \lbrack 6,7]~\mathrm{GeV}^{2},\ s_{0}^{\prime }\in \lbrack
35,36]~\mathrm{GeV}^{2},
\end{equation}%
for $\overline{B}_{s}^{\ast 0}$ channel. The constants in the fit function $%
\widehat{\mathcal{F}}_{1}(Q^{2})$ are $\widehat{\mathcal{F}}_{1}^{0}=0.0119~%
\mathrm{GeV}^{-1}$, $\widehat{l}_{1}^{1}=13.11$, and $\widehat{l}%
_{1}^{2}=-16.36$. The corresponding strong coupling $\widehat{G}_{1}$ is
equal to
\begin{equation}
\widehat{G}_{1}=(9.75\pm 1.87)\times 10^{-3}\ \mathrm{GeV}^{-1}.
\end{equation}%
Then, the partial width of the decay $\mathcal{M}_{\mathrm{AV}}\rightarrow
\overline{B}_{s}^{\ast 0}D_{s}^{-}$ amounts to

\begin{equation}
\Gamma \left[ \mathcal{M}_{\mathrm{AV}}\rightarrow \overline{B}_{s}^{\ast
0}D_{s}^{-}\right] =(4.3\pm 1.2)~\mathrm{MeV}.
\end{equation}

The next channel with the final-state strange mesons $\mathcal{M}_{\mathrm{AV%
}}\rightarrow \overline{B}_{s}^{0}D_{s}^{\ast -}$ is treated in the context
of similar analyses. Our prediction for the coupling $\widehat{G}_{2}$ reads
\begin{equation}
\widehat{G}_{2}=(3.21\pm 0.67)\times 10^{-3}\ \mathrm{GeV}^{-1}.
\end{equation}%
Then, the partial width of the decay $\mathcal{M}_{\mathrm{AV}}\rightarrow
\overline{B}_{s}^{0}D_{s}^{\ast -}$ equals to

\begin{equation}
\Gamma \left[ \mathcal{M}_{\mathrm{AV}}\rightarrow \overline{B}%
_{s}^{0}D_{s}^{\ast -}\right] =(2.6\pm 0.8)~\mathrm{MeV}.
\end{equation}%
Note that this result has been obtained utilizing the parameters
\begin{equation}
M_{2}^{2}\in \lbrack 5.5,6.5]~\mathrm{GeV}^{2},\ s_{0}^{\prime }\in \lbrack
34,35]~\mathrm{GeV}^{2},
\end{equation}%
in $\overline{B}_{s}^{0}$ meson's channel. The coupling $\widehat{G}_{2}$
has been estimated by means of the extrapolating function $\widehat{\mathcal{%
F}}_{2}(Q^{2})$ with $\widehat{\mathcal{F}}_{2}^{0}=0.004~\mathrm{GeV}^{-1}$%
, $\widehat{l}_{2}^{1}=12.78$, and $\widehat{l}_{2}^{2}=-16.64$.

By taking into account all results obtained in this and previous sections,
it is easy to evaluate the full decay width of the hadronic molecule $%
\mathcal{M}_{\mathrm{AV}}$:
\begin{equation}
\Gamma \left[ \mathcal{M}_{\mathrm{AV}}\right] =(114\pm 17)~\mathrm{MeV}.
\end{equation}%
This result shows that $\mathcal{M}_{\mathrm{AV}}$ is a rather broad
compound, where the subleading decays are essential and form up to $30\%$ of
its full width.


\section{Analysis and Final Notes}

\label{sec:Conc}


In the present article, we have continued our studies of the exotic hadronic
molecular structures with asymmetric quark contents. In Ref. \cite%
{Agaev:2025did}, we explored the scalar states $\eta _{b}B_{c}^{-}$ and $%
\eta _{c}B_{c}^{-}$ with contents $bb\overline{b}\overline{c}$ and $cc%
\overline{c}\overline{b}$, respectively. In this work, we have considered
the axial-vector structures $\mathcal{M}_{\mathrm{AV}}=\Upsilon B_{c}^{-}$
and $\widetilde{\mathcal{M}}_{\mathrm{AV}}=\eta _{b}B_{c}^{\ast -}$ by
calculating their masses and decay widths. It turned out that these
particles have, at least within accuracy of the SR method, identical masses.
Therefore, we have assumed that their decay widths are approximately equal
to each other and studied in detail decay channels of $\mathcal{M}_{\mathrm{%
AV}}$.

Our prediction $m=(15800\pm 90)~\mathrm{MeV}$ for the mass of the molecule $%
\mathcal{M}_{\mathrm{AV}}$ means that it is strong-interaction unstable
particle and decays to $\Upsilon B_{c}^{-}$ and $\eta _{b}B_{c}^{\ast -}$
pairs of ordinary heavy mesons. Apart from these two dominant modes, there
are channels of $\mathcal{M}_{\mathrm{AV}}$ which become possible due to
annihilation of $b\overline{b}$ quarks in $\mathcal{M}_{\mathrm{AV}}$ and
generation of light quarks' pairs. We have investigated decays to $B^{\ast -}%
\overline{D}^{0}$, $\overline{B}^{\ast 0}D^{-}$, $B^{-}\overline{D}^{\ast 0}$%
,$\overline{B}^{0}D^{\ast -}$, $\overline{B}_{s}^{\ast 0}D_{s}^{-}$, and $%
\overline{B}_{s}^{0}D_{s}^{\ast -}$ mesons from the class of subleading
processes.

The fully heavy hadronic molecules were object of interesting analyses in
Ref.\ \cite{Liu:2024pio}, in which the authors applied the extended local
gauge formalism to investigate such systems with different spin-parities.
The masses of the hadronic molecules were estimated there using the cutoff
momentum $\Lambda $ of the model. The authors made interesting conclusions
about properties of the fully heavy axial-vector $\Upsilon B_{c}^{-}$, $\eta
_{b}B_{c}^{\ast -}$ and tensor $\Upsilon B_{c}^{\ast -}$ molecules. In
accordance with their predictions, these structures reside below the
corresponding two-meson thresholds and form bound states. The mass of the
axial-vector molecules $\Upsilon B_{c}^{-}$ and $\eta _{b}B_{c}^{\ast -}$
was found there equal to $15725.3~\mathrm{MeV}$ at $\Lambda =1000~\mathrm{MeV%
}$.

This result is smaller than our prediction provided one compares it with the
central value $m=15800~\mathrm{MeV}$, but at the lower limit, $m=15710~%
\mathrm{MeV}$, they are comparable with each other. The hadronic molecule
with the mass $15710~\mathrm{MeV}$, of course, can not decay to $\Upsilon
B_{c}^{-}$ and $\eta _{b}B_{c}^{\ast -}$ mesons. Nevertheless the subleading
channels for its transformation to conventional particles remain open is
this case as well. To elucidate this question we have computed the width of $%
\mathcal{M}_{\mathrm{AV}}$ with the mass $15710~\mathrm{MeV}$. Analysis has
been carried out in accordance with the scheme presented in section \ref%
{sec:Widths2}. Our predictions for the partial decay widths of these modes
are collected in Table\ \ref{tab:Channels}. The width of the molecule $%
\mathcal{M}_{\mathrm{AV}}$ in this case is $\Gamma \left[ \mathcal{M}_{%
\mathrm{AV}}\right] =(30\pm 4)~\mathrm{MeV}$, which is not small and
determines the mean lifetime of this structure.

\begin{table}[tbp]
\begin{tabular}{|c|c|c|}
\hline\hline
Channels & Strong c.$\times 10^{3}~(\mathrm{GeV}^{-1})$ & $\Gamma~(\mathrm{%
MeV})$ \\ \hline
$B^{\ast -}\overline{D}^{0}$ & $13.2 \pm 2.8$ & $7.9 \pm 2.4$ \\
$B^{-}\overline{D}^{\ast 0}$ & $3.76 \pm 0.75$ & $3.9 \pm 1.1$ \\
$\overline{B}^{\ast 0}_{s}D^{-}_{s}$ & $9.34 \pm 1.75$ & $3.8 \pm 1.0 $ \\
$\overline{B}_{s}^{0}D_{s}^{\ast -}$ & $3.07 \pm 0.62$ & $2.3 \pm 0.7 $ \\
\hline\hline
\end{tabular}%
\caption{Decay modes of the molecule $\mathcal{M}_{\mathrm{AV}}$ at the
lower limit of its mass $15710~\mathrm{MeV}$, relevant strong couplings, and
partial widths $\Gamma$.}
\label{tab:Channels}
\end{table}

Studies performed in present article has important consequences for
understanding features of the hadronic molecules $\Upsilon B_{c}^{-}$ and $%
\eta _{b}B_{c}^{\ast -}$. It has been demonstrated that heavy mesons $%
\Upsilon $ and $B_{c}^{-}$ (as well as, $\eta _{b}$ and $B_{c}^{\ast -}$)
may form a bound molecular state which does not dissociate to constituent
particles. But due to annihilation of $b\overline{b}$ quarks it is unstable
structure and has relatively small full width. Alternatively, the mesons $%
\Upsilon $ and $B_{c}^{-}$ may establish a broad compound. The first
scenario is consistent with findings of Ref.\ \cite{Liu:2024pio}. Evidently,
these theoretical problems should be confirmed in the context of presumably
alternative models and methods.

Obtained results are also important for experimental investigations of fully
heavy structures. They provide information about masses and possible decay
channels of hadronic molecules $\mathcal{M}_{\mathrm{AV}}$ and $\widetilde{%
\mathcal{M}}_{\mathrm{AV}}$ pointing out final-states where they may be
discovered. Thus, observation of an enhancement (a peak) in the mass
distributions of the meson pairs $B^{\ast -}\overline{D}^{0}$, $B^{-}%
\overline{D}^{\ast 0}$ and others from Table\ \ref{tab:Channels} would be
considered as evidence for realizations of the first scenario for $\mathcal{M%
}_{\mathrm{AV}}$, whereas the similar effect for mesons $\Upsilon B_{c}^{-}$
interpreted as correctness of the second option. Production mechanisms of
heavy molecules $\mathcal{M}_{\mathrm{AV}}$ and $\widetilde{\mathcal{M}}_{%
\mathrm{AV}}$ are interesting problems as well. But their investigations
require independent and detailed investigations which are beyond the scope
of the current work.

\end{document}